# THERMOEMF OF POWDER-BASED THERMOELECTRIC MATERIALS

*Gorskyi P.V.*

*Institute of Thermoelectricity of the NAS and MES of Ukraine, 1 Nauky str., Chernivtsi, 58029, Ukraine*

*In the paper, the thermoEMF of powder-based thermoelectric materials (TEM) is calculated. The calculation is made on the assumption of power dependence of mean free path on energy. The thermoEMF decreases with increasing the average radius of powder particles, however, it drastically increases with an increase in power exponent in the law of dependence of the mean free path of relaxation time on energy (scattering index). Therefore, it turns out that the thermoEMF of powder-based TEM with a higher scattering index can be even greater than the thermoEMF of a single-crystal material with a low scattering index. As a consequence, a significant increase in the thermoEMF and, hence, in the thermoelectric figure of merit of TEM in going to powder materials, especially in the case of degenerate electron gas, can be expected only if dielectric or vacuum barriers between powder particles do not lead to a significant decrease in electrical conductivity. At the same time, tunnelling through the abovementioned barriers should provide such an energy filtration of charge carriers, which leads both to an increase in the proportion of "useful" charge carriers with energy greater than the chemical potential, and to an increase in the scattering index. However, experimentally, there is no significant increase in thermoEMF in going from single-crystal materials to powders, most likely because such energy filtering does not take place.*

**Introduction**

It is considered that the thermoelectric figure of merit of powder-based TEM should increase considerably, since this is facilitated by energy barriers occurring due to vacuum or dielectric gaps between powder particles [1-3]. Tunnelling of free charge carriers through these gaps should increase the contribution of charge carriers with high energies to kinetic coefficients, including thermoEMF. If this filtering is strong enough, i.e. charge carriers that have passed through and not passed through the barriers differ significantly in energy, then it should lead to a significant increase in thermoEMF for two reasons. First, if the electron gas is degenerate, albeit weakly, then an increase in the fraction of charge carriers with an energy greater than the chemical potential should lead to a significant increase in the thermoEMF due to the fact that the contact and volume components of the thermoEMF are opposite in sign. In the nondegenerate gas, on the contrary, the signs of these components are identical, the contact component being the main one. Therefore, in the nondegenerate gas this factor plays a lesser role. Second, according to the general principles of quantum mechanics, the scattering index $r$ must increase as the energy of the charge carriers increases [4]. If we are dealing with a material for which the quadratic law of charge carrier dispersion is valid, then the inequality $-0.5 \leq r \leq 3.5$ is true for this index. In semiconductor TEM, in the area relevant for thermoelectric applications, the situation is most often realized, in which the value $r$ is the smallest possible and corresponds to scattering of charge carriers with energy-independent cross-section, hence, the mean free path length [5]. In this case the thermo EMF of a powder-based material does not depend on the particle size of the powder, and therefore is the same as that of a single-crystal material, if we do not consider the possibility of changing the character of energy spectrum of charge carriers, for instance, due to dimensional quantization at particularly small particle sizes [4]. However, this parameter can be increased, in particular, by alloying the TEM [6]. Therefore, the purpose of this paper is to consider the dependence of the thermoEMF on the powder particle size arising when the scattering index increases beyond the minimum value.

**General formula for the thermoEMF of powder-based TEM**

To derive a general formula determining the thermoEMF of a powder-based TEM, we will use the approach we developed earlier in [7]. We will proceed from the general formula for the thermoEMF in the impurity conduction region, which in the case of an isotropic TEM with a parabolic law of charge carrier dispersion can be represented as

$$\alpha = \frac{\int_0^\infty v^2(\varepsilon)\tau(\varepsilon)\varepsilon\left(-\frac{\partial f^0}{\partial \varepsilon}\right)g(\varepsilon)d\varepsilon}{eT\int_0^\infty v^2(\varepsilon)\tau(\varepsilon)\left(-\frac{\partial f^0}{\partial \varepsilon}\right)g(\varepsilon)d\varepsilon} - \frac{\zeta}{eT}. \qquad (1)$$

In this formula: $e$ – electron (hole) charge with corresponding sign, $T$ – absolute temperature, $\varepsilon$ – electron (hole) energy, $f^0$ – the Fermi-Dirac equilibrium distribution function, $v(\varepsilon)$, $\tau(\varepsilon)$, $g(\varepsilon)$ – energy-dependent only electron velocity, their relaxation time and density of states, respectively, $\zeta$ – the chemical potential of the TEM electronic subsystem.

The relaxation time $\tau(\varepsilon)$ in our case depends on both the scattering mechanisms characteristic of single crystals and the scattering of electrons (holes) at the boundaries of powder particles. As a consequence, it can be represented as:

$$\tau(\varepsilon) = \left\langle \frac{\tau_0 \varepsilon^r L}{\tau_0 \varepsilon^r v(\varepsilon) + L} \right\rangle. \qquad (2)$$

In this formula: $\tau_0$ – some constant value with respect to energy, $L$ – possible mean free path of electron due to scattering at the boundaries of the powder particle, angular brackets denote averaging over all possible free path lengths within the particle.

In turn, the constant $\tau_0$ can be expressed in terms of the mean free path of an electron (hole) $l_{e,h}$ in a single-crystal material, based on the ratio:

$$l_{e,h} = \frac{A\tau_{0e,h}\int_0^\infty \varepsilon^{r+1} f^0(\varepsilon)d\varepsilon}{\int_0^\infty \sqrt{\varepsilon} f^0(\varepsilon)d\varepsilon} = A\tau_{0e,h}(kT)^{r+0.5}\frac{F_{r+1}(\eta)}{F_{1/2}(\eta)}. \qquad (3)$$

In this formula: $A$ – the coefficient of proportionality between the velocity of electron and the square root of its energy, $F_m(\eta)$ – the Fermi index integral $m$, $\eta = \zeta/kT$.

By determining from (3) $\tau_{0e,h}$, substituting the result in (2), averaging over the free path lengths inside the powder particle, which, for simplicity, we assume as a sphere of radius $r_0$, and finally calculating the thermoEMF using formula (1), we arrive at the following result:

$$\alpha = \frac{k}{e}\left\{\frac{\int\limits_0^\infty\int\limits_0^1\int\limits_{-1}^1 f_1(x,\eta)x^{r+2.5}\left[\frac{x^{r+0.5}F_{1/2}(\eta)}{F_{r+1}(\eta)} + \frac{r_0}{l_{e,h}}\sqrt{y^2+1+2zy}\right]^{-1} y^2 dzdydx}{\int\limits_0^\infty\int\limits_0^1\int\limits_{-1}^1 f_1(x,\eta)x^{r+1.5}\left[\frac{x^{r+0.5}F_{1/2}(\eta)}{F_{r+1}(\eta)} + \frac{r_0}{l_{e,h}}\sqrt{y^2+1+2zy}\right]^{-1} y^2 dzdydx} - \eta\right\}. \quad (4)$$

In this formula, $f_1(x,\eta)-$ the energy derivative of the Fermi-Dirac distribution function, other notations explained above or generally accepted.

It is easy to see from this formula that, first, at $r=-0.5$ the thermoEMF is independent of the powder particle radius, second, at $r_0/l_{e,h} \to \infty$ the thermoEMF tends to the value characteristic of single crystals.

**Results of calculating the thermoEMF of powder-based TEM and their discussion**

Results of calculating the thermoEMF of powder-based TEM by formula (4) are given in Fig.1.

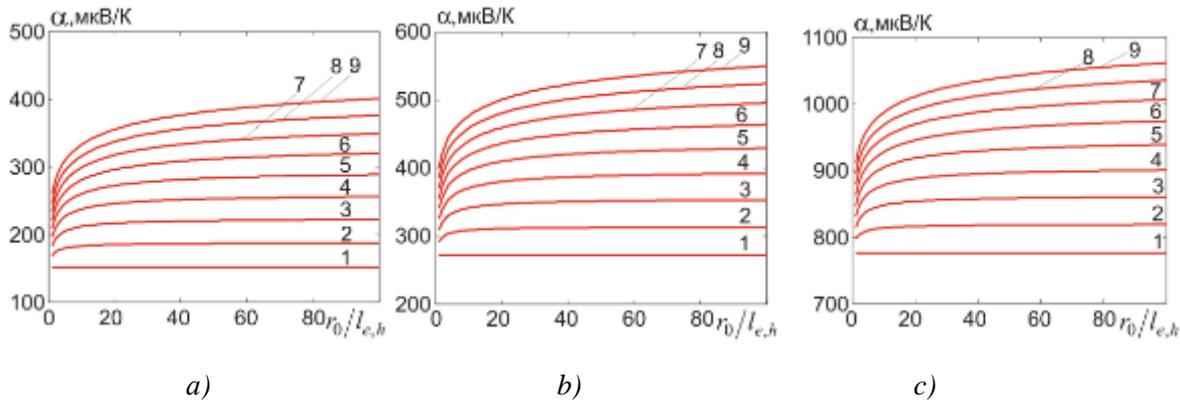

*a)*          *b)*          *c)*

*Fig.1. Dependences of the thermoEMF of powder-based TEM on the radius of powder particle at: a) $\eta=1$; b) $\eta=-1$; c) $\eta=-7$. In each figure, curves 1-9 are plotted for scattering index $r$ values between –0.5 and 3.5 in 0.5 increments.*

The figures show that for all considered degrees of degeneracy at the minimum possible scattering index, the thermoEMF for a single crystal and powder-based TEM is the same and has a minimum value. However, if the scattering index differs from the minimum possible, the thermoEMF becomes dependent on the size of the powder particle and decreases in magnitude as its radius decreases, the greater the scattering index. However, if we assume that the scattering index does not change in going from a single crystal to powder, then even at its maximum value a tangible decrease in thermoEMF occurs only at particle radii less than 40 electron (hole) free path lengths. Taking into account that the typical value of this length for materials based on the *Bi(Sb)-Te(Se)* system at 300K is about 20 nm, it turns out that a significant decrease in thermoEMF can occur only in materials based on submicron or nanopowders with an average particle radius of less than 0.8 μm.

It should be noted, however, that the assumption that the scattering index is unchanged in going from a single crystal to powder is somewhat simplified. The point is that the grinding of a single-crystal material to a powdery state and its subsequent fabrication as a TEM by hot pressing, extrusion, or electrospark plasma sintering involves a number of mechanical and thermal effects on the initial single crystal, which affect the defectiveness of the material, and, consequently, the scattering mechanisms of charge carriers in it, and thus the value of the scattering index. If the starting material is manufactured in such a way that in the temperature range of application the scattering index is greater than the minimum value, there are factors that can both increase and decrease the value of the scattering index in going from a single crystal to powder. The formation of vacuum or dielectric barriers in the gaps between the TEM particles, provided that their height and width are optimal for changing the scattering index with a satisfactory probability of charge carriers passing through the barriers, should be considered an enhancing factor. The formation of structural defects inside the material particles with an energy-independent or weakly energy-dependent scattering cross-section of charge carriers on them should be considered a decreasing factor. Thus, on the one hand, the possibility of a significant increase in the figure of merit of TEM in going from a single crystal to powder is determined by the ratio between the competing processes of energy filtration of charge carriers and formation of defects with an energy-independent or weakly energy-dependent cross section of scattering of charge carriers on them. On the other hand, it follows from our earlier studies [7] that in going from a single crystal to powder, the electrical conductivity decreases at any value of the scattering index, and much more strongly than the thermoEMF. Thus, in going from a single crystal to powder with small particle size, the power factor of the material decreases, owing to which the thermoelectric figure of merit of the material can increase only with a sharp decrease in thermal conductivity, mainly due to its lattice (phonon) component. However, in a powder material, the thermoEMF can be fully or partially recovered by annealing, which reduces the number of defects with an energy-independent or weakly energy-dependent cross section of scattering of charge carriers on them. And such recovery does occur in some thermoelectric materials based, for example, on submicron powders [8]. But this mechanism alone still cannot fully explain the change in thermoEMF in going from single crystals to powders. The fact is that, for example, according to the data of [8] the ratio of electrical conductivity to thermal conductivity increased by a factor of 100 in going from a single crystal to powder with an average particle radius equal to 0.2 μm, while the figure of merit increased only 6-fold, which indicates a decrease of thermoEMF by approximately 4 times. And even when the scattering index drops from maximum to minimum, the thermoEMF decreases by a factor of only 2.7 at the degrees of degeneracy considered in this paper. Thus, either the charge carrier gas was more strongly degenerate, or there are other mechanisms that lead to thermoEMF decrease.

**Conclusions and recommendations**

1. It is shown that in powder-based materials the thermoEMF decreases with decreasing average powder particle radius the more the relaxation time depends on the energy. However, at particle radii larger than the free path of the charge carriers, it is still larger than even in a single crystal, provided that the free path of the charge carriers is energy-independent.

2. A significant decrease in the thermoEMF in going from a single crystal to a powder-based material can only occur if either the scattering index decreases due to the formation of defects that scatter charge carriers with an energy-independent cross section or if a dimensional quantization of the energy spectrum of charge carriers occurs as the size of the powder particle decreases.

3. The thermoEMF of powder-based TEM can be restored completely or, at least partially, as a result of annealing, reducing the concentration of "harmful" defects with a small scattering index.

The author is grateful to Acad. L.I. Anatychuk for his approval of the research topic and to L.N. Vikhor, chief researcher, for helpful and constructive discussion of the results of the work.